# Room-temperature Photoluminescence from $Er^{3+}$ in Si-Er-O and Si-Ge-Er-O Thin Films at High Erbium Concentrations

S. Abedrabbo [1,2*] and A. T. Fiory [2]

[1] *Department of Physics, University of Jordan, Amman, Jordan 11942*
[2] *Department of Physics, New Jersey Institute of Technology, Newark, NJ 07901 USA*

## Abstract

Prior studies have shown that photoluminescence from $Er^{3+}$ impurities in silicon is severely limited at room temperature by non-radiative relaxation and solid solubility; and room temperature emission from $Er^{3+}$ in oxide-based hosts becomes diminished at high erbium concentrations. This work presents studies of thin films (0.2 μm thick) prepared by vacuum co-evaporation from elemental sources (Er, Si and Si/Ge) followed by vacuum annealing (600°C); materials of this type, which are produced with high $Er^{3+}$ concentrations, are shown to be capable of yielding strong room-temperature photoluminescence. Alloy films of Si-Er-O and Si-Ge-Er-O, containing (20 ± 2) at. % Er and incorporating (16 ± 2) at. % O (introduced via vacuum scavenging reactions), exhibit emission bands with dominant components at 1.51 and 1.54 μm (~0.04-μm overall spectral widths). Results are discussed in terms of Er-O complex formation and effects of local randomness on cooperative inter-$Er^{3+}$ energy transfer among thermal-broadened and local-field Stark-split $^4I_{13/2} \rightarrow {}^4I_{15/2}$ transitions. Advantages of scalability and low-cost associated with this method of producing optically active silicon-based materials are discussed.

*Corresponding author: sxa0215@yahoo.com; sxa0215@njit.edu

# 1. Introduction

Increasing attention is being focused on optically active $Er^{3+}$ centers in Si, owing to intra-4$f$ luminescence in the 1.54-μm region, for a broad range of applications, e.g. light emitters that are integrated with silicon-based optoelectronics (Kenyon, 2005, and references therein). Optical emission from Er in crystalline Si is known to suffer from thermal quenching (e.g. reduced emission at room temperature), which can be ameliorated to some extent by oxygen co-doping (Kenyon, 2005; Coffa *et al.*, 1994). In contrast to crystalline Si, $Er^{3+}$ is optically active at room temperature in various non-crystalline hosts, notably in $SiO_2$ and silica-based dielectric materials for Er-doped glass-fiber amplifiers (Desurvire, 1994). Well-studied approaches for circumventing thermal quenching are the Er-doped sub-stoichiometric Si-rich oxide films, e.g. $SiO_{2-x}$:Er (Kenyon *et al.*, 1994; Lenz *et al.*, 2009), in which optical emission from $Er^{3+}$ is enhanced relative to Er-doped $SiO_2$, owing to proximity of $Er^{3+}$ ions to semiconductor nanocrystal inclusions embedded in the oxide host (Fujii *et al.*, 1998; Gourbilleau *et al.*, 2003; Chen *et al.*, 2003).

Previous works reported room-temperature photoluminescence (RT-PL) from $Er^{3+}$ in Si films alloyed with high Er concentrations and co-doped with oxygen (Abedrabbo *et al.*, 2009, 2010). Viability of using a semiconductor-based approach was demonstrated for the purpose of reducing thermal quenching by the formation of O-containing alloys as well as quenching associated with high Er concentrations (dissipative Er-Er interactions) by virtue of local randomness within the host material. Doping with oxygen (a high electronegativity ligand) modifies the local environment around Er atoms (low electronegativity and high affinity to O), increases Er solubility and inhibits Er segregation by forming Er-O complexes (Coffa *et al.*, 1994; Michel *et al.*, 1998; Adler *et al.*, 1992). This present work is an extended study of RT-PL spectra that compares results obtained for Si-Ge-Er-O and Si-Er-O thin films; importance of these semiconductor-alloy based thin-film materials derives from their capability of generating RT-PL, in contrast to the weak or near absence of emission at room temperature that is characteristic of crystalline hosts at low Er concentration (including co-doping with O, such as obtained for Czochralski-grown silicon material) (Kenyon, 2005; Coffa *et al.*, 1994). Oxygen is introduced at concentrations (14 – 18 at. % O) that considerably exceed solid solubility limits in bulk-silicon crystals and yet are well below the O content of $SiO_{2-x}$:Er films ($x \sim 10^{-2}$). Several observed systematic behaviours are of interest. Emission is found to be more intense for a Si-Ge-Er-O alloy film, when compared to Si-Er-O containing comparable levels of oxygen and erbium content, although the $Er^{3+}$ emits at nearly equivalent dominant emission wavelengths (Stark-split $^4I_{13/2} \rightarrow {}^4I_{15/2}$ transitions near 1.51 and 1.54 μm wave lengths) for the two materials. An interpretation invoking cooperative energy transfer among dipolar coupled optically active $Er^{3+}$ ions in the presence of alloy randomness is introduced to assist in understanding these observations, in particular the apparent suppression of concentration quenching at high Er concentrations (~20 at. %).

Interest in materials containing erbium in high concentrations (i.e. exceeding the equilibrium solubility in bulk-silicon crystals) stems also from recent observations of RT-PL in 0.2-μm silica films heavily doped with Er (6 at.% Er, produced by a low-cost sol-gel spin-coating deposition and thermal annealing process) (Abedrabbo *et al.*, 2011a, 2011b). Relatively high Er concentrations have also been successfully incorporated in $SiO_2$ waveguide structures that exhibit high gains and intense PL signals (Park *et al.*, 2003). Potential benefits of Ge co-doping can be inferred from related studies of Er in crystalline and amorphous hosts. Originally, Ge was incorporated in silica fiber cores containing Er as a network modifier and a refractive index enhancer (Desurvire, 1994). Emission from Er in Si-Ge quantum wells is notably stronger at low temperature than in Si (Evans-Freeman *et al.*, 2001). In a more closely related study, Ge co-doping in silica was found to produce enhanced RT-PL from Er and was attributed to the formation of nascent Ge nanostructures (Heng *et al.*, 2004). The above considerations therefore inspired the present strategy of combining disordered Si/Ge structures, high Er concentrations, and incorporation of sufficient O concentration.

## 2. Experiment

Thin-film (0.2 µm in thickness) specimens were prepared by co-evaporation in vacuum ($10^{-4}$ Pa) of elemental Si, Ge and Er onto single-crystal Si held at 300 °C. The substrates were 1-cm square sections cleaved from double-side polished Si wafers ($\langle 100 \rangle$ orientation, boron doped at $10^{15}$ cm$^{-3}$ concentration, and 250 µm in thickness) containing native-oxide surfaces. After deposition, films were annealed for 1 hr. in vacuum at 600 °C, which is in the optimum range for RT-PL from Er$^{3+}$ in oxide-based hosts (400 °C − 800 °C) (Chen *et al.*, 2003; Heng *et al.*, 2004; van den Hoven *et al.*, 1995; Franzò *et al.*, 2003; Ha *et al.*, 2003) and was found to favor Er-O complex formation and decrease the density of non-radiative decay centers (Adler *et al.*, 1992; Franzò *et al.*, 2003); the annealing temperature is also low enough to avoid phase separation and large scale nano-crystal growth (Heng *et al.*, 2004). Oxygen is incorporated by uptake from the evaporants and substrate, which contain native oxides, and residual gas in the vacuum chamber; this technique was demonstrated earlier in the deposition of crystalline (Stepikhova *et al.*, 2000) and amorphous (Terukov *et al.*, 1998) Si films. Inclusion of Ge with moderately low annealing temperature is thus expected to reduce the sizes of semiconductor nanocrystals, in a manner similar to the observations of deposited (Si/Ge)O$_{2-x}$:Er films by Heng *et al.* (2004) for Ge and Er doped silica, which in that case yielded enhanced PL emission by tailoring the band gap to suppress exciton dissociation (typical films produced for sub-oxide materials studies are ~ 1-µm thick).

Central points addressed in this work therefore concern the roles of (1) high Er concentration and (2) Ge co-doping (i.e. its substitution for Si). The thin-film samples examined for this study contain Er, Ge, and O in various proportions and yield RT-PL emission from the Er$^{3+}$. Silicon-germanium alloyed with erbium, which was studied for several films of average Ge concentrations of 22 − 65 %, was found to be capable of producing an enhanced RT-PL relative to silicon-erbium alloys without germanium.

The effect of Ge co-doping was studied by comparing in particular two samples based on Si-Ge-Er and Si-Er alloys and containing comparable levels of erbium as well as oxygen impurities. A specimen denoted as sample 1 was prepared with a 170-nm Si-Ge-Er-O film deposited in two stages: a 70-nm under layer film was deposited by co-evaporation of Si and Er and a 100-nm over layer was deposited on top by co-evaporation of Si, Ge, and Er. A second specimen (sample 2) was prepared with a 200-nm Si-Er-O film by co-deposition of Si and Er (i.e. no Ge). The purpose of the under layer structure in sample 1 was to maintain the same interface constituents at the surface of the Si substrate for both specimen types (thereby removing Si substrate interactions as another potential materials variable).

Film compositions (Si, Ge, Er, and O content) and thicknesses were determined by Rutherford backscattering spectrometry (RBS) by modelling the samples as multiple layers, in which species concentrations and layer thicknesses were varied for best fit (for RBS analysis methodology, see Abedrabbo *et al*., 2009, 2010). Film thicknesses were obtained by interpolating atomic densities of Si, Er, Ge, and their oxides. Steady-state RT-PL spectra (0.4-s integration time per datum) were obtained with a model Flourolog-3 spectrofluorometer (courtesy of Horiba Jobin-Yvon, Edison, New Jersey (Horiba Jobin-Yvon, 2007)) using an excitation wavelength of 530 nm (Hg-lamp source) and a liquid-N$_2$-cooled InGaAs photodiode (Hamamatsu). The spectrofluorometer is a self-calibrating instrument with all of the optics being reflective to circumvent the distortions induced by lenses. Neodymium-doped phosphate laser glass (emission peak at $1.051 \pm 0.0005$ μm) serving as a reference standard was used to independently check the calibration of the InGaAs detector and the coupling optics.

## 3. Results

Figure 1 (a) shows distributions of Ge and Er concentrations (averaging 46% Ge and 19% Er, respectively, in the top 100 nm of the film) in the Si-Ge-Er-O sample 1; while Figure 1 (b) shows the Er distribution (averaging 22% Er) in the Si-Er-O sample 2; these concentration profiles were determined by RBS analysis. The Er distributions in the active top (near-surface) 100-nm of the two samples are quite similar to one another: the Er doping constitutes an atomic area density of $0.80 \times 10^{17}$ cm$^{-2}$ in the top 100 nm in sample 1 and $0.98 \times 10^{17}$ cm$^{-2}$ in the top 100 nm of sample 2.

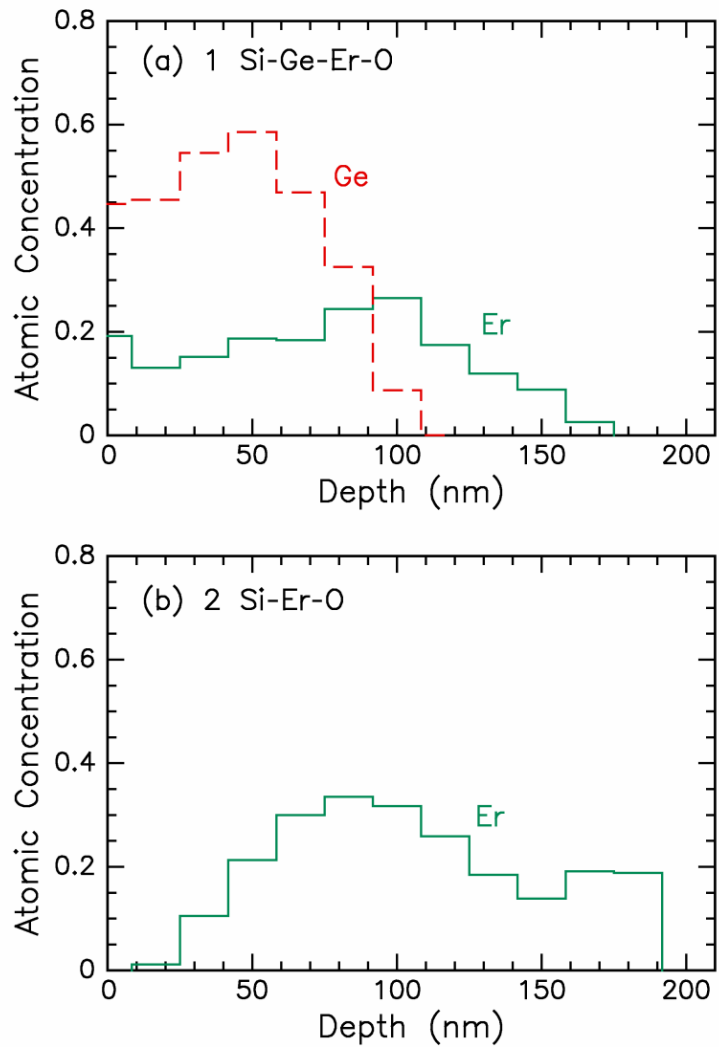

**Figure 1**. Depth dependences of atomic concentrations from RBS analysis. (a) Ge and Er in sample 1, Si-Ge-Er-O, and (b) Er in sample 2, Si-Er-O.

Sample 2 has the larger total area density of Er ($1.20\times10^{17}$ cm$^{-2}$ in sample 1; $1.64\times10^{17}$ cm$^{-2}$ in sample 2). Both samples contain significant amounts of O as well, particularly in the top 100-nm layer, where the oxygen concentration is estimated to be 14% for sample 1 and 18% for sample 2. Thus the oxygen content is substantially less than in the more oxygen-rich films employed in previous studies of erbium doping in sub-stoichiometric oxides, e.g. $SiO_{2-x}$ and $(Si-Ge)O_{2-x}$ (Heng *et al.*, 2004).

While sample 2 may contain a greater quantity of potentially optically active $Er^{3+}$ (i.e. measurably higher amounts of Er and O), the emission intensities turn out to trend in the opposite manner. Figure 2 shows the steady-state RT-PL for the two sample types, clearly indicating that the PL signal from the Si-Ge-Er-O sample 1 is more intense than that from the Si-Er-O sample 2. With similar thicknesses for the active layers and comparable Er distributions in the two films (i.e., the top 100 nm of both samples), the important difference in composition appears to be the substitution of Ge for a portion of the Si. Both spectra exhibit two readily resolved peaks near 1.510 – 1.511 μm and 1.538 μm (luminescence at 0.977 μm, corresponding to de-excitation from the next highest $^4I_{11/2}$ levels, is also observed). The spectra shown in Figure 2 are similar in form to PLspectra observed for $Er^{3+}$ in silica and $SiO_{2-x}$ suboxides containing Si nanostructures (Fujii *et al.*, 1998; Gourbilleau *et al.*, 2003; Chen *et al.*, 2003; Franzò *et al.*, 2003; Stepikhova *et al.*, 2000) and oxygen-doped silicon (van den Hoven *et al.*, 1995). The well-resolved double peak spectra are reminiscent of emission from non-alumina-based optical fibers containing Er impurities and with or without incorporation of $GeO_2$ (Ge concentrations at much lower levels than in the present study) (Kenyon *et al.*, 1994).

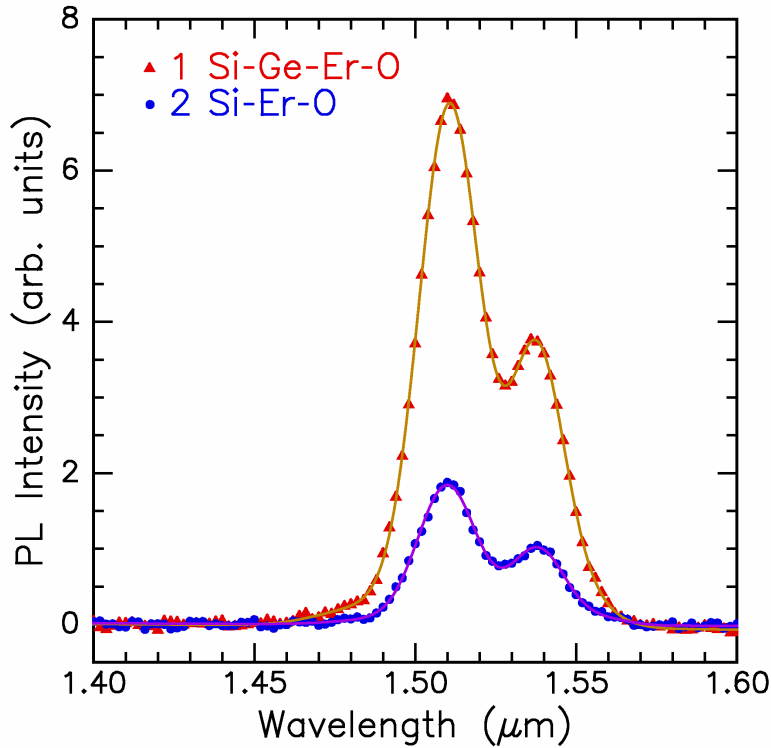

**Figure 2**. Photoluminescence spectra (points) at room temperature of sample 1, Si-Ge-Er-O (red triangles) and sample 2, Si-Er-O (blue circles). Curves are fitted with Gaussian components (parameters in Table 1) for sample 1 (brown) and sample 2 (purple).

**Table 1.** Multiple-Gaussian fit parameters for PL spectra of samples 1 (Si-Ge-Er-O) and 2 (Si-Er-O) shown in Figure 2. Widths are Gaussian standard deviations.

| Sample | Wavelength (μm) | Intensity (arb. units) | Width (μm) |
|---|---|---|---|
| 1 Si-Ge-Er-O | 1.486 | 0.25 | 0.015 |
|  | 1.511 | 6.85 | 0.010 |
|  | 1.538 | 3.48 | 0.008 |
|  | 1.550 | 0.38 | 0.010 |
| 2 Si-Er-O | 1.489 | 0.06 | 0.012 |
|  | 1.510 | 1.83 | 0.009 |
|  | 1.538 | 1.01 | 0.009 |
|  | 1.559 | 0.08 | 0.007 |

The spectra in Figure 2, in which full width at half maximum (FWHM) is 42 nm in both cases, are understood in terms of inhomogeneous broadening of Stark-split $^4I_{13/2} - {}^4I_{15/2}$ transitions in $Er^{3+}$, since the optical activity of $Er^{3+}$ behaves similarly as in optical fibers (Kenyon *et al.*, 1994).

One finds that a superposition of four Gaussian functions, with two dominant components and two weaker satellite components, is able to provide a good fit to the overall spectral line shapes, as shown by the solid curves in Figure 2. The fitted components (central wavelengths, amplitudes in terms of the arbitrary-units scale of the PL spectra, and widths as given by the standard deviation of the Gaussian function) are presented in Table 1. On comparing areas under the Gaussian-resolved spectral components, emission peaking at 1.510 — 1.511 μm is enhanced (Si-Ge-Er-O relative to Si-Er-O) by a factor of 3.7 and at 1538 nm by a factor of 3.5. Comparing total areas under the two data spectra, Er emission from sample 1 is enhanced by factor of 3.8, relative to sample 2; the higher integrated emission intensity from sample 1 owes to the relatively stronger (> 4× greater) emission in the wings of the spectra (satellite components).

## 4. Discussion

The unique attribute of doping with Ge concerns the ability to produce statistically random mixtures (i.e. absence of preferential clustering) that are characteristic of equilibrium Si-Ge alloys. In the case of bulk Si-Ge crystals, the intrinsic statistical fluctuations of random mixing is known to promote the binding and localizalization of excitons (Weber and Alonzo, 1989); further, Ge excitons are thought to enhance recombination excitation of the $Er^{3+}$ 4-*f* levels (Heng *et al.*, 2004). Silicon-germanium alloying also increases site-to-site disorder, which is expected to enhance carrier recombination at Er-O defects and create the inhomogeneous broadening that is desirable for wavelength selection in optoelectronic devices (Desurvire, 1994).

Inhomogeneous spectral widths for $Er^{3+}$ emission from the Si-Ge-Er-O and Si-Er-O films are comparable to findings for Er in oxide- and silicon-based hosts (e.g. 38 – 45 nm) (Abedrabbo *et* *al.*, 2009; Heng *et al.*, 2004; van den Hoven *et al.*, 1995). Site symmetry variations in disordered structures are known to produce inhomogeneous broadening of the Stark-split energy sublevels in the ground $I_{15/2}$ and first excited $I_{13/2}$ manifolds by as much as 25 cm$^{-1}$ (equivalent to ~6-nm broadening of $^4I_{13/2} - {^4I_{15/2}}$ transitions) (Desurvire, 1994). Thus, given that the films are disordered and contain substantial oxygen, it is fair to draw comparisons with Er-containing glassy networks, where co-doping reduces Er microscopic clustering at high concentration by forming a shell around the rare-earth ion, thus facilitating its dispersal within the matrix (Zemon *et al.*, 1991). We deem this to be an important consideration in both samples, and particularly for the Si/Ge sample 1. Details of the emission peaks exhibited in RT-PL spectra are determined by site symmetries and thermal distributions of occupied Stark-split sublevels and are therefore materials dependent. In film materials with high Er concentrations, emission from $^4I_{13/2} - {^4I_{15/2}}$ transitions of Er$^{3+}$ ions have been observed at various wavelengths; in erbium-silicate films (1000 – 1200 °C anneals), the dominant emission wavelengths are near 1.50, 1.53, and 1.55 μm (Miratello *et al.*, 2007; see RT-PL spectra in Figure 3a therein); in spin-coated sol-gel films (containing 6% Er; ~ 750 – 900 °C anneals) the dominant emission is near 1.537 μm (Abedrabbo *et al.*, 2011a); in Si-Er-O films, ion beam mixing (post 600-°C annealing) was found to change the dominant emission wavelength from 1.516 to 1.534 μm (Abedrabbo *et al.*, 2009). (A preliminary study of ion-beam mixing of Si-Ge-Er-O indicates RT-PL with maximum at 1.529 μm and 40 nm FWHM.)

Optically active Er$^{3+}$ is generally associated with a symmetry-reducing ligand dipole (e.g., an ErO$_n$ cluster with n ≈ 6) (Kenyon *et al.*, 1994; Michel *et al.*, 1998; Heng *et al.*, 2004). For the films of the present study, the probability of a single Er-O coordination may be assumed to scale with the ratio $p$ of O to Er concentrations, owing to the high electronegativity difference of Er-O relative to Si-O or Ge-O and to the high Er concentration. This follows from observations that RT-PL tends to increase with oxygen content: Figure 3 shows the integrated RT-PL intensity for a series of Si-Er-O samples (samples 3 − 5 are from Abedrabbo *et al.*, 2009) plotted against the ratio of area densities of oxygen ($N_O$) and erbium ($N_{Er}$) as determined from RBS analysis. The fraction of Er in randomly formed $ErO_n$ structures with O coordination number n ($1 \leq n \leq 6$) is estimated by the expression $F_n = p^n(1-p)/(1-p^6)$. For samples 1 and 2, we have $p \sim 0.7$ and thus derive the result $F_6 = 0.04$ (a conservative underestimate, since this approximation does not account for O sharing in Er-O bonding). Given that the O/Er fraction (Figure 3) and Er concentration (Figure 1) are, for samples 1 and 2, comparable to one another, the implication is that the increased RT-PL emission observed for sample 1 (see Figure 2) can be associated with the presence of Ge in that alloy (sample 1 has optimal Ge/Er content among the several Si-Ge-Er-O samples studied).

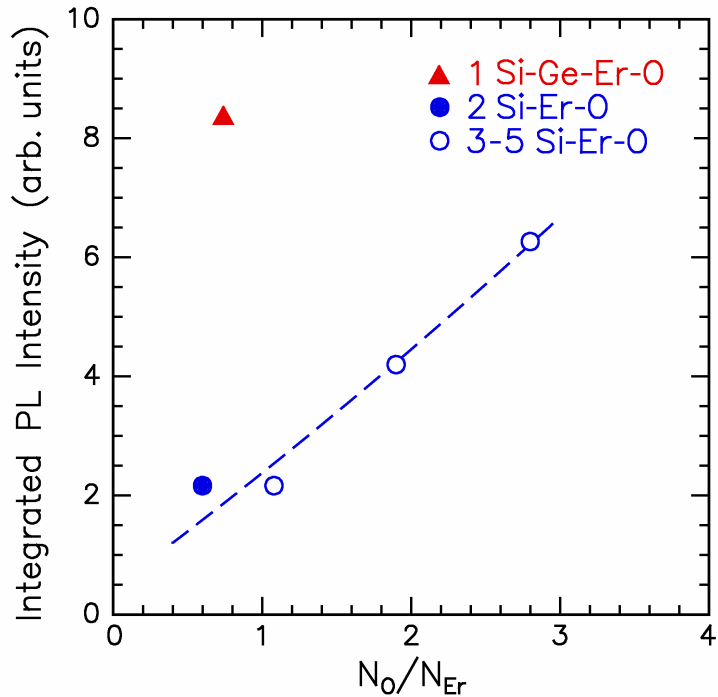

**Figure 3.** Integrated photoluminescence intensity (1.4 μm – 1.6 μm) *vs.* ratio of area atomic densities of oxygen and erbium (RBS analysis) for Si-Ge-Er-O (sample 1, filled red triangle) and Si-Er-O (sample 2, filled blue circle; samples 3-5, open blue circles). Dashed line denotes trend.

Since the above estimates indicate a substantial density (possibly $\sim 10^{21}$ cm$^{-3}$) of $Er^{3+}$ ions in optically active environments (i.e. in Er-O complexes), the influence of concentration quenching effects are to be expected; these involve the transfer of the excitation energy among the optically active $Er^{3+}$ by dipole-dipole interactions when inter-ion separations are within ~1 nm. One such mechanism is associated with fast diffusion of the excitation energy via the Förster mechanism, commonly referred to as cooperative energy transfer (CET), where $Er^{3+}$ de-excitation energy is non-radiatively dissipated at defects or traps (Desurvire, 1994). Related modes of inter-ion interaction are the cooperative up-conversion (CUP) processes in which two units of excitation energy is dissipated by doubled-energy excitation of an $Er^{3+}$ ion (Wyatt, 1990; Desurvire, 1994). In this work, we note that RT-PL is observed in the Si-Er-O and Si-Ge-Er-O films, despite the expectation of concentration quenching effects; this suggests that the alloy randomness shields some $Er^{3+}$ sites from the dissipative Er-Er interactions that would otherwise occur at high Er concentrations (i.e. two dominant sites indicated by the results in figure 2 and table 1). Following this line of reasoning, one would expect a greater degree of randomness in a quaternary material (Si-Ge-Er-O), as compared to a ternary material (Si-Er-O); this provides a natural explanation for the differences between samples 1 and 2. This picture of concentration quenching, i.e. being suppressed by local disorder, is consistent with a previous interpretation of RT-PL observed in Er-doped silica films containing high Er concentrations (Abedrabbo *et al.*, 2011a).

A moderate thermal annealing temperature of 600 °C is used in this work in order to favor formation of Er-O complexes and to disfavor detrimental phase separation. Low-temperature annealing is also desirable for integration with silicon microelectronics. High-temperature annealing, on the other hand, is expected to increase thermal quenching, as previously demonstrated for Er-doped $SiO_{2-x}$ films, where it was attributed to nano-crystal formation and energy back transfer (Lenz *et al.*, 2009). High temperature annealing is especially deleterious for $GeO_{2-x}$ hosts, where RT-PL of $Er^{+3}$ disappears upon precipitation of Ge nano-crystals (Ardyanian *et al.*, 2007). Moderate annealing temperatures have also been found to be sufficient for decreasing defects (which may trap Er emission radiation) created during deposition (van den Hoven *et al.*, 1995; Ha *et al.*, 2003). In Er and Ge co-doped silica, RT-PL emission is strongest for annealing at 700 °C and drops markedly for high-temperature annealing (> 800 °C) as Ge nano-crystals are nucleated and grow (as determined by x-ray diffraction analysis) (Heng *et al.*, 2004). Of particular significance is that the annealing temperature for strongest $Er^{+3}$ emission lies below the threshold of discernable nano-crystal formation. Hence, the selection of 600 °C as the annealing temperature in this work is considered to be optimum for enhancing Er-O formation, decreasing nonradiative decay centers (van den Hoven *et al.*, 1995; Ha *et al.*, 2003), while also minimizing the formation of sizable nano-crystals.

Energy transfer from electron-hole pairs excited by the pump radiation to defect centers associated with Er-O complexes is a plausible interpretation of observed excitation of $Er^{3+}$, e.g. by trapped or localized recombining electron-hole pairs or excitons (Michel *et al.*, 1998), given the high Er concentrations. In light of the present PL results, the instructive conclusion from prior work on $Er^{+3}$ in suboxides and oxides, amorphous-like Si-structures and Si-Ge alloys is that local disorder in the host is favorable for suppressing thermal quenching in optically active $ErO_6$-like structures (Marcus and Polman, 1991; Adler *et al.*, 1992). The limiting case of maximum disorder includes Er-doped hydrogenated amorphous Si films (Terukov *et al.*, 1998; Bresler *et al.*, 1995; Zanatta and Freire, 2000), where RT-PL is interpreted as energy transfer of excited carriers in Urbach tail states with impurity centers associated with the Er-O complexes (Janotta *et al.*, 2003). While Er emission is found to be enhanced in Si-Ge alloys, thermal quenching effects remain a significant impediment for device applications (Ishiyama *et al.*, 2006; Stepikhova *et al.*, 2006; Kamiura *et al.*, 2008; López-Luna *et al.*, 2009). Nevertheless, alloying disorder introduces several important properties in Si-Ge systems. In amorphous Si-Ge, the widths of exponential band tails (primarily the conduction band) increase with Ge concentration (Wang *et al.*, 1993; Aljishi *et al.*, 1990). Alloying has the known effect of inducing local fluctuations in the band edges (Petrosyan and Shik, 1982) and promoting formation of bound and localized excitons that are observed by photoluminescence in equilibrium Si-Ge alloys (Weber and Alonzo, 1989) and Si-Ge superlattices (Ghosh *et al.*, 2000). In their recent study of $SiO_2$ containing nano-crystalline Si and Er, Fujii et al. reported that energy transfer to Er is facilitated by the distributed exciton recombination rates in the Si nano-crystals, owing to band gap variations with nano-crystal size (Fujii *et al.*, 2004).

Improving $Er^{+3}$ emission therefore entails incorporating O to form Er-O defects and alloying Si with Ge to stabilize statistical randomness on the atomic scale. Further, high Er concentrations increase the density of optically active Er while also promoting random alloy mixing. By using a cost-efficient processing method with moderate vacuum levels, the present results point to effectiveness of high Ge concentration in producing RT-PL in a Si-Ge-Er-O layer of $1/5^{th}$ the typical thickness used for sub-oxide materials. The Ge alloying is assumed to depress nonradiative decay channels while also creating better energy transfer for exciting the $Er^{3+}$ ions by direct and indirect means.

Given that the formation of quaternary materials involves numerous process variables, more detailed studies are needed to determine the best process parameters, and the optimum concentrations of Ge, Er and O for device structures emitting in the 1.54-μm band.

## 5. Conclusions

Observations of room-temperature photoluminescence from $Er^{3+}$ have shown the viability of generating optically activity at high Er/O concentrations (~20% Er, ~16 %O) from ternary (Si-Er-O) and quaternary (Si-Ge-Er-O) thin film (~ 0.2 μm thick) materials produced by low-cost vacuum deposition from pure elemental sources and thermal annealing at a low thermal budget (600 °C). These studies have demonstrated that in principle, such films merit interest as emerging materials suitable for silicon-chip based optoelectronics applications, where high Er concentrations are advantageous, owing to the limited space for optical interaction available for fully on-chip planar devices (e.g. when compared to optical-fiber devices containing $Er^{3+}$ at low concentrations). The results suggest that optical activity of $Er^{3+}$ can be improved with Si/Ge alloying. Oxygen incorporation plays an essential role, in that the optically active $Er^{3+}$ centers are associated with Er-O structures.

It would be of interest to design further experimentation to evaluate the interrelationships among the various process variables that were presently touched upon. These include (1) Er concentrations and profiles, (2) Ge/Si ratios and profiles, (3) controlled introduction of oxygen and other elemental co-dopants (e.g. Yb, N, C), (4) thermal annealing (temperatures, times, ambients), and (5) multiply layered structures (e.g. inclusion of passivation and buffer layers). Also of interest is engaging further microanalysis techniques for determining film structures and compositions (e.g. x-ray diffraction, electron microprobe, transmission electron microscopy, x-ray photoemission spectroscopy); and expanded study of optical activity (e.g. photoemission excitation matrices, photoemission lifetimes, and electro-optical structures). The film deposition and annealing methods utilized in the process described in this work are readily scalable to large areas, rapid through-put and low cost of ownership in production; further, the materials and processes involved are compatible (the low total thermal budget in particular) for integrated silicon-based planar optoelectronics devices (active and passive components).

## Acknowledgements

The authors are grateful to the Deanship of of Academic Research at the University of Jordan, Horiba Jobin-Yvon and the New Jersey Institute of Technology for partly funding their project, to the staff of University of Jordan for their support, and to F. Abu Sa'n. It is indeed a pleasure to acknowledge the support and hospitality provided by Prof. N. M. Ravindra. Publication on this work has appeared (Abedrabbo, 2011c).